\documentclass[3p,times,procedia]{elsarticle}
\usepackage{nupha_ecrc}

\volume{00}

\firstpage{1}

\journalname{Nuclear Physics A}

\runauth{G.A. Alm\'{a}si et al.}

\jid{nupha}

\jnltitlelogo{Nuclear Physics A}

\usepackage{amssymb}
\usepackage{hyperref}

\usepackage[figuresright]{rotating}

\begin{document}

\begin{frontmatter}


\title{Modeling chiral criticality and its consequences for heavy-ion collisions}


\address[GSI]{Gesellschaft f\"{u}r Schwerionenforschung, GSI, D-64291 Darmstadt, Germany}
\address[EMMI]{ExtreMe Matter Institute (EMMI), D-64291 Darmstadt, Germany}
\address[Wroclaw]{University of Wroc\l aw - Faculty of Physics and Astronomy, PL-50-204 Wroc\l aw, Poland}
\address[Duke]{Department of Physics, Duke University, Durham, NC 27708, USA}

\author[GSI]{G\'abor Andr\'as Alm\'asi}
\ead{g.almasi@gsi.de}
\author[GSI,EMMI]{Bengt Friman}
\ead{b.friman@gsi.de}
\author[EMMI,Wroclaw,Duke]{Krzysztof  Redlich}
\ead{krzysztof.redlich@ift.uni.wroc.pl}

\dochead{}

\begin{abstract}

We explore the critical fluctuations near the chiral critical
endpoint (CEP) in a chiral effective model
and discuss possible signals of the CEP, recently explored  experimentally in nuclear
collision. Particular attention is paid to the dependence of
such signals on the location of the phase boundary and the CEP relative to the chemical freeze-out conditions in nuclear collisions. We argue that in effective models, standard freeze-out fits to heavy-ion data should not be used directly. Instead, the relevant quantities should be examined on lines in the phase diagram that are defined self-consistently, within the framework of the model. We discuss possible choices for such an approach.
\end{abstract}


\end{frontmatter}


\section{Introduction}
\label{sec:intro}
The chiral phase transition of QCD is conceptually well understood in terms of effective models, that belong to the same universality class. However, a quantitiative characterization of this transition is at present available only at small net baryon densities. Thus, it is well established, that at low values of the baryon chemical potential, the system undergoes a smooth crossover transition with increasing temperature from a phase, where the chiral symmetry is spontaneously broken to a phase, where the chiral symmetry is approximately restored~\cite{Aoki:2006we}. It has been conjectured that at high net baryon densities, the transition could be  first order. In this case the first order transition line would end in a critical point which is referred to as the Critical Endpoint of QCD (CEP)~\cite{Asakawa:1989bq,Barducci:1989wi,Wilczek:1992sf,Halasz:1998qr}. In spite of intense efforts both in theory and experiment to establish the nature of the chiral transition at non-zero baryon density, the possible existense of a chiral critical endpoint remains unclear.

One of the goals of the Beam Energy Scan (BES) program at the Relativistic Heavy Ion Collider (RHIC) is to explore QCD phase diagram at non-zero net baryon density by probing fluctuations of conserved charges along the chemical freeze-out line~\cite{Stephanov:1998dy,Stephanov:1999zu,Asakawa:2000wh,Jeon:2000wg,Friman:2011pf}. In the following we will focus on fluctuations of the net baryon number, which are reflected in the measured cumulants of the net proton event-by-event fluctuations~\cite{Adamczyk:2013dal}.

We present calculations of the baryon number cumulants
	$\chi^n_B = T^{n-4}\partial^n P(T,\mu_B)/\partial \mu_B^n$
 from the pressure $P$ in the function of temperature $T$ and baryon chemical potential $\mu_B$ in the Polyakov--Quark-Meson (PQM) model within the framework of the Functional Renormalization Group (FRG)~\cite{Wetterich:1992yh,Morris:1993qb,Berges:2000ew}. By changing the parameters of the model we are able to modify the location of the CEP. We determine the freeze-out line in the phase diagram self-consistently for each parameter set of the model and compute the baryon number cumulants along these lines. In such a scheme, part of the model dependence is expected to be eliminated. We discuss the relationship between various baryon number cumulants and compare our findings with the recent data obtained in the BES~\cite{Adamczyk:2013dal}.

\section{The Polyakov--Quark-Meson model}
\label{sec:PQM}

The Polyakov--Quark-Meson (PQM) model~\cite{Schaefer:2007pw} is an effective theory of QCD, which describes chiral symmetry breaking and simulates confinement of quarks at low temperatures and chemical potentials and free quark degrees of freedom at high temperatures. These properties allow us to study the interplay of the chiral and deconfinement phase transitions in this model. The degrees of freedom of the model are quarks (up and down in this work) and mesons (the pions and their chiral partner, the scalar sigma meson). Furthermore, the quarks are coupled to a background gluon field which manifests itself in the Polyakov loop. Details of the model can be found in the literature~\cite{Schaefer:2007pw,Skokov:2010sf}.


The fluctuations of both quark and meson fields are accounted for using the functional renormalization group. This is done by solving the flow equation~\cite{Wetterich:1992yh}, which evolves the effective average action by successively integrating out fluctuations of a decreasing momentum scale, starting from the classical action in the ultraviolet to the full quantum effective action in the infrared. We work in the Local Potential Approximation (LPA), where only the mesonic potential is scale dependent. We discretize the order parameter space and solve the flow equation on a grid~\cite{Herbst:2013ail}.

\section{Results}

The interpretation of the baryon number cumulants obtained in heavy-ion collisions is a non-trivial task. In addition to the difficulties involved in assessing finite size and finite time effects, there are several sources of non-critical fluctuations~\cite{Kitazawa:2011wh,Skokov:2012ds,Bzdak:2013pha},which are not yet fully understood. Moreover, in a comparison with effective models, one has to deal with the fact that non-universal quantities can differ
from their values in QCD. This means, e.g., that the pseudo-critical temperature of the model will in general be different from that of QCD. Owing to this, a direct comparison between model and experiment could be very misleading. One can reduce this model dependence either by a suitable rescaling of the freeze-out line or by a self-consistent determination of this line within the model. We consider the following three possibilities to locate the freeze-out line:
\begin{itemize}
	\item heavy ion freeze-out line taken from \cite{Andronic:2008gu}, with $T_0$ fitted to the pseudo-critical temperature at $\mu = 0$ for all parameters,
	\item pseudo-critical line,
	\item $\chi^3_B / \chi^1_B = 0.9$ line.
\end{itemize}

This procedure approximately preserves the relation between the freeze-out line and the chiral transition, thus eliminating part of the model dependence. The motivation for defining the third line emanates from the experimental data: recent STAR results \cite{Adamczyk:2013dal} show that in the energy range of the measurements the ratio $\chi^3_B / \chi^1_B$ remains approximately constant at a value slightly below unity. In the hadron resonance gas (HRG) this ratio is unity, while in a gas of free quarks it would be $6/\pi^2$. Consequently, a reduction of this ratio is a conseuqence of the change in degrees of freedom. It follows from the universal structure of the singular free energy near the CEP that all lines of constant $\chi^3_B / \chi^1_B$ converge on and pass through the CEP. The CEP is a singular point, where the cumulant ratio $\chi^3_B / \chi^1_B$ takes any value between $-\infty$ and $+\infty$, depending the path of approach. Our results are almost independ of the exact value chosen for $\chi^3_B / \chi^1_B$, as long as it lies in the interval $[6/\pi^2,1]$.

At small net baryon densities, the lines are close together and yield similar results for the cumulant ratios. On the other hand, for large densities the first line ends up in the hadronic phase, while the second and the third lines pass through the CEP. Consequently, the resulting values for the cumulant ratios are very different.

\begin{figure}[tb]
   \includegraphics[width=0.49\linewidth]{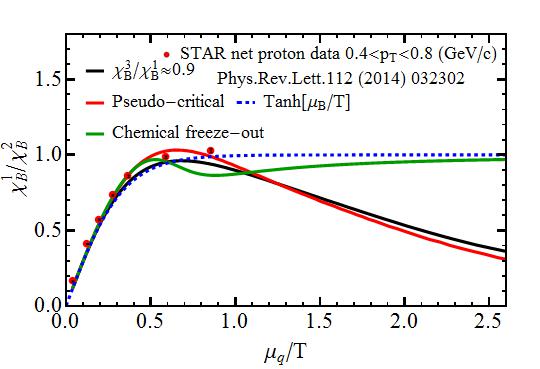}
   \includegraphics[width=0.49\linewidth]{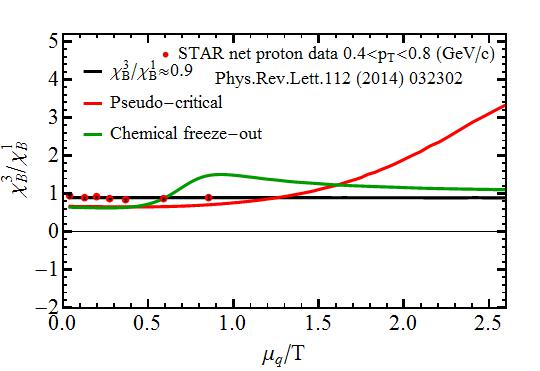}
   \includegraphics[width=0.49\linewidth]{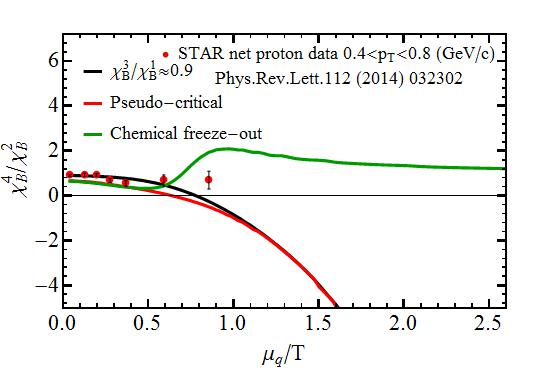}
   \hspace{0.12 cm}
   \includegraphics[width=0.49\linewidth]{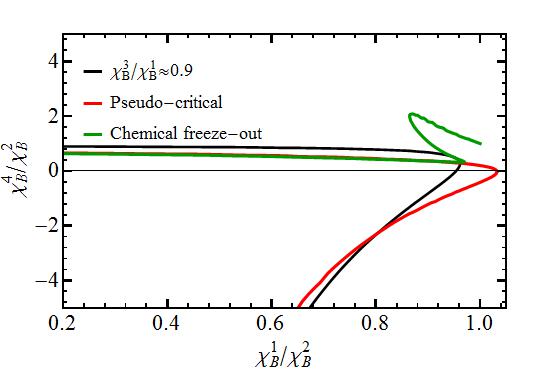}
   \caption{Cumulant ratios on different line of the phase diagrams. More details can be found in the main text.\label{fig:cumulratios}}
\end{figure}

As shown in the left top panel of Fig. \ref{fig:cumulratios}, the experimental $\chi^1_B / \chi^2_B$ cumulant ratio is very well reproduced by the HRG result $\tanh\left(\mu_B/T\right)$ and reasonably well by the model for all choices of the freeze-out line. On the other hand, the model results for $\chi^3_B / \chi^1_B$ and $\chi^4_B / \chi^2_B$, shown in the top right and bottom left panels, clearly depend on the freeze-out line already in the range of the experimental data. Moreover, both ratios clearly deviate from the HRG prediction, i.e., from unity. At higher values of the chemical potential, as the CEP is approached, these deviations are enhanced. On the pseudo-critical and the $\chi^3_B / \chi^1_B = 0.9$ lines, the ratio $\chi^1_B / \chi^2_B$ decreases and eventually vanishes at the CEP, while  $\chi^4_B / \chi^2_B$ decreases on both lines, and approaches $-\infty$. 
 We also note that on the pseudo-critical line, the ratio $\chi^3_B / \chi^1_B$ grows and diverges at the CEP.
The above properties of cumulants of the net baryon number can be understood following discussions presented in Refs.  ~\cite{Friman:2011pf,Karsch:2005,Bengt:2011,Stephanov:2011pb}.


At large values of $\mu_q/T$, the results obtained on the (rescaled) phenomenological freeze-out line differ substantially from the other ones. This is related to the corresponding freeze-out conditions. For large values of the chemical potential, this line ends up in the hadronic phase, well below the phase boundary obtained in the model. Consequently, on this line all cumulant ratios approach unity at large values of $\mu_q/T$. Thus the general structure of the singular free energy near the CEP implies that both ratios $\chi^3_B / \chi^1_B$ and $\chi^4_B / \chi^2_B$ are enhanced along this freeze-out line at large values of the chemical potential.

In the bottom right panel of Fig. \ref{fig:cumulratios} we show the cumulant ratio $\chi^4_B / \chi^2_B$ as a function of $\chi^1_B / \chi^2_B$. Depending on the freeze-out line, a characteristic non-monotonous behaviour is found as the CEP is approached. In this way of plotting the results, a direct comparison with data is possible, eliminating part of the model dependence~\cite{Karsch:2015nqx}.

\section{Conclusions}

We presented results for ratios of baryon number cumulants obtained in a chiral effective model using the FRG to include fermionic and bosonic fluctuations. We found that on the $\chi^3_B / \chi^1_B = const$ lines, $\chi^4_B / \chi^2_B$ decreases as the CEP is approached, while on a freeze-out line where $\chi^4_B / \chi^2_B$ increases, $\chi^3_B / \chi^1_B$ also grows, starting at approximately the same value of the chemical potential. In light of these results, it is unclear how the recent preliminary STAR data on net baryon number fluctuations~\cite{Luo:2015ewa, Luo:2015doi, Thader:2016gpa} can be interpreted. These data exhibit an enhancement of $\chi^4_B / \chi^2_B$ at low beam energies, while $\chi^3_B / \chi^1_B$ remains almost constant.

\section*{Acknowledgments}
The work of B.F. and K.R.  was partly supported by the Extreme Matter Institute EMMI. K. R. also  acknowledges  partial  supports of the Polish Science Foundation (NCN) under the Maestro grant DEC-2013/10/A/ST2/00106, and the  U.S. Department  of  Energy  under  Grant  No.  DE-FG02-05ER41367. G. A. acknowledges the support of the Hessian LOEWE initiative
through the Helmholtz International Center for FAIR (HIC for FAIR).





\bibliographystyle{elsarticle-num}
\bibliography{refs}






\end{document}